\documentclass[fleqn,10pt]{wlscirep}
\usepackage[utf8]{inputenc}
\usepackage[T1]{fontenc}
\usepackage{dsfont}
\usepackage{amsmath,scalerel}

\usepackage{url, hyperref}

\usepackage[norelsize, noend, linesnumbered, ruled, lined, boxed, commentsnumbered]{algorithm2e}

\title{Hashing for Secure Optical Information Compression in a Heterogeneous Convolutional Neural Network}

\author[1]{Maria Solyanik-Gorgone}
\author[1, 2]{Behrouz Movahhed}
\author[1, 2, *]{Volker J Sorger}
\affil[1]{Department of Electrical and Computer Engineering, George Washington University, Washington DC, 20052}
\affil[2]{Optelligence LLC, 10703 Marlboro Pike, Upper Marlboro, MD, 20772, USA}
\affil[*]{sorger(@gwu.edu, @optelligence.co)}

\begin{abstract}
In the recent years, heterogeneous machine learning accelerators have become of significant interest in science, engineering and industry. The major processing speed bottlenecks in these platforms come from (a) an electronic data interconnect; (b) an electro-optical interface update rate. In this light, information compression implemented in native to incoming data optical domain could mitigate both problems mentioned above by reducing the demand on data throughput at the camera side and beyond. In this paper we present an optical hashing and compression scheme that is based on SWIFFT  - a post-quantum hashing family of algorithms. High degree optical hardware-to-algorithm homomorphism allows to optimally harvest well-understood potential of free-space processing: innate parallelism, low latency tensor by-element multiplication and Fourier transform. The algorithm can provide several orders of magnitude increase in processing speed by replacing slow high-resolution CMOS cameras with ultra fast and signal-triggered CMOS detector arrays. Additionally, the information acquired in this way will require much lower transmission throughput, less \emph{in silico} processing power, storage, and will be pre-hashed facilitating cheap optical information security. This technology has a potential to allow heterogeneous convolutional 4f classifiers get closer in performance to their fully electronic counterparts.
\end{abstract}
\begin{document}

\flushbottom
\maketitle
\thispagestyle{empty}

%\noindent Please note: Abbreviations should be introduced at the first mention in the main text – no abbreviations lists. Suggested structure of main text (not enforced) is provided below.

\section{Introduction}

Optical compression methods have been extensively studied for decades inspired by bandwidth, storage and processing power hungry real-time image recognition and holography \cite{naughton2003efficient, dufaux2015compression, zhang2020optical}. Recently, the transformative advances in technology such as internet of things, edge-computing and 3D visualisation with light-field cameras posed new challenges, such as algorithms capable of real-time image pre-processing, compression and pattern recognition that resulted into the emergence of a new field of compressive sensing \cite{donoho2006compressed, qaisar2013compressive}. Large image size and strong temporal correlations that distinguish multimedia data from other formats pose a great computational challenge due to convolutions and dot-product multiplications being the main component of image processing ML \cite{li2016performance, jorda2019performance}, and simultaneously the main computer power consumer. 

In most currently deployed heterogeneous implementations, optical data processing, encryption and compression are accomplished electronically \cite{naughton2002compression, mills2005effects, singh2010chaos}, hence, \emph{after} the electro-optical domain conversion. Often, image quality reduction is implemented first to alleviate corresponding processing load. This way, the benefits of having a high-resolution camera in free-space setting are not fully harvested due to the digital image processing in order to meet electronic streaming throughput and/or data storage limitations. Alongside with that, other venues for development recently have been gaining momentum such as pre-processing in photonic integrated circuits \cite{juleang2021optical}, custom CMOS chips for optical encryption \cite{zhou2022photonic}, and extra-low formfactor metasurface-based optical systems \cite{zhou2019optical}. These shows multi-directional and cross-disciplinary interest in novel hardware solutions for high-speed data processing and security.

Explored here, Fourier optics based heterogeneous data compression and encryption is by far not a new idea \cite{situ2003cascaded, hennelly2003optical, situ2004double, jin2007optical, lang2010image, zhou2011novel, zhang2018multiple}. Free-space 4f-systems have been previously considered for optical encryption in phase-only modulating setups, oftentimes employing random phase filters \cite{situ2003cascaded, jin2007optical}. Exotic proposals include approximating digital fractional Fourier transform algorithms in Fourier optics \cite{lang2010image}. However, to our knowledge, there has been no attempt to (1) directly adopt post-quantum encryption protocols in the field of optical processing; (3) simulate simultaneous amplitude-and-phase modulation filtering in Fourier domain to achieve confusion step in heterogeneous hashing; and (3) achieve simultaneous optical hashing and compression.

Fourier 4f-system-based \cite{goodman2005introduction} simultaneous optical input hashing and compression system presented here elegantly addresses multiple drawbacks of optical processing: (a) feature extraction; (b) feature hashing; (c) data compression; (d) data security. As opposed to compact and fast on-chip solutions \cite{juleang2021optical}, throughput of the featured optical compression system is limited only by the resolution of the light modulating device and diffraction properties of light, while phase control can be achieved via conventional free-space alignment methods. The theoretical core of the design is based on one of the lattice-based post-quantum hashing algorithms \cite{ajtai1996generating} and uses it as a prototype for the featured here implementation of heterogeneous 4f-based optical data hashing. SWIFFT family of FFT-based hashing functions considered here are not lossless, but asymptotically secure, highly parallelizable, collision resistant and algorithmically homomorphic to optical processing \cite{lyubashevsky2008swifft}. The underlying class of linear algebra operations for lattice-based cryptography, matrix-vector multiplication, has shown the potential to be performed optically \cite{miscuglio2021massively} or in a photonic chip \cite{miscuglio2020photonic} with projected efficiency exceeding currently existing fully-electronic accelerators, supplied high-speed domain conversion and electronic interconnect. With a thin lens being a passive Fourier transform engine \cite{goodman2005introduction}, the algorithm has a potential to elegantly integrate into modern optical interconnect potentially providing an unprecedentedly fast additional layer of security.

The proposed optical setup can be realized on a set of spatial light modulators (SLMs) with employed full amplitude-phase convolution in Fourier domain of a 4f system. This ASIC optical co-processor is capable of hashing 2D arrays of binary data at kHz speed and a fundamental latency in optical domain of $1.49 \cdot 10^8$s, limited by the speed of light in free space. Key practical limitations come from the update rate of (1) modulating devices; (2) image acquisition and processing hardware; (3) speed of the supporting electronic interconnect. First limitation, high-speed SLMs with kHz update rate that could be an optimal choice for this project, remain to be a major holdup in the field of optical processing currently limiting the achievable speed of heterogeneous co-processors by 120Hz (off-the-shelf). However, if compression speed is a priority as opposed to hash security, the requirement on SLMs' update rate can be substantially relaxed. In the limiting case, SLMs can even be replaced with static masks not compromising compression quality. Second and third limitations can be mitigated by an order of magnitude in speed due to the one-dimensional optical output of the proposed scheme. That allows one to use ultra-high-speed sensor arrays (e.g. $\sim0.5$MHz Linea ML detector arrays) as opposed to many earlier proposed architectures that require order of magnitude slower CCD/CMOS cameras at the domain crossing and ultra-high bandwidth electronic interconnect. The back-end processing can be accomplished via comparing to a hash-table or using ML algorithms depending on the optical system's and information channel's stability, and a particular application. 
%hector (IARPA) - 10^9->10^6.

\begin{figure}[t!]
\begin{center}
\includegraphics[width=0.8\linewidth]{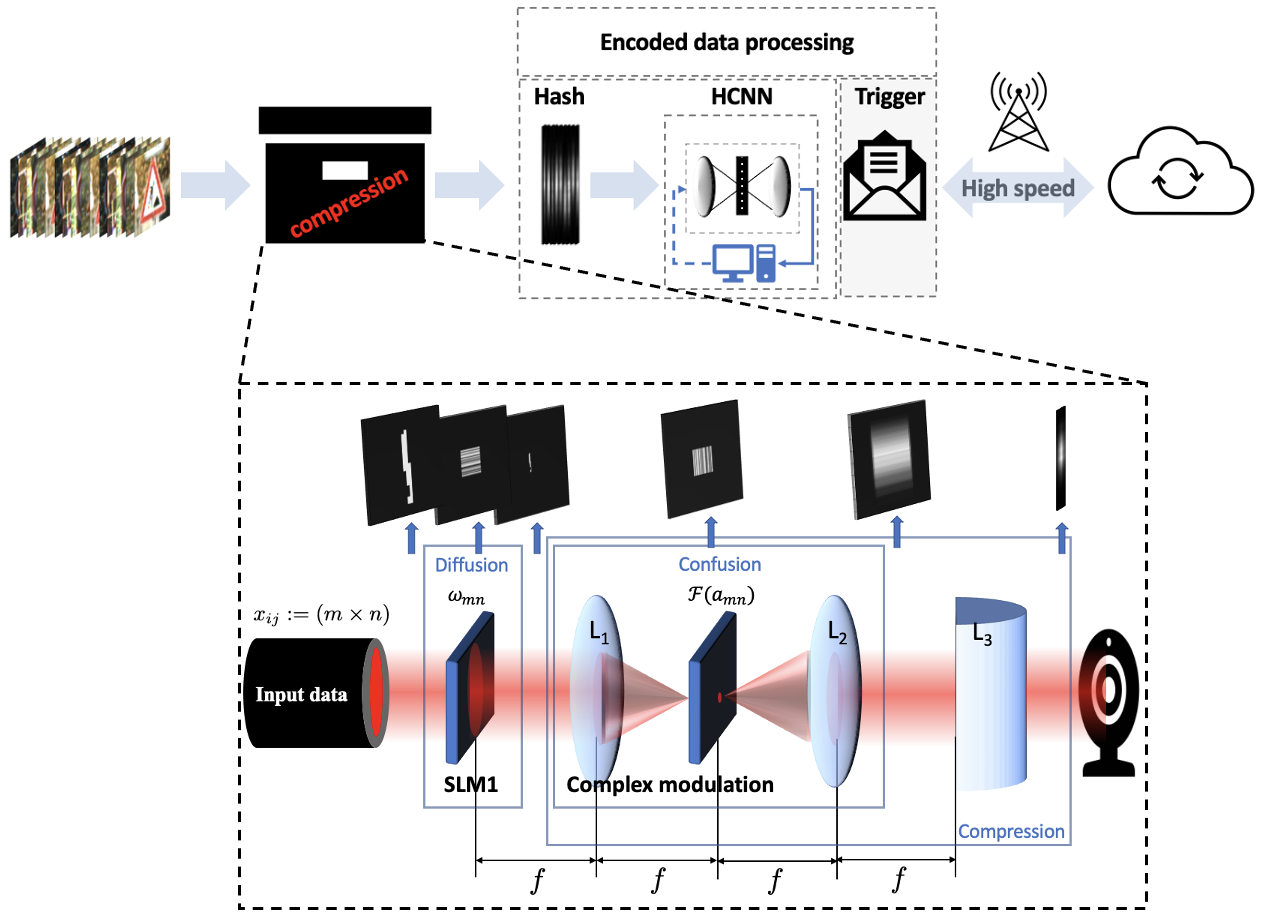}
\caption{The concept of the SWIFFT-based heterogeneous (optical) hashing/compression system and its integration into a generic edge system data interconnect. The advantages of (1) harvesting the features of the input in analogue domain, hence, before introducing analogue-to-digital conversion related noise; (2) utilising low-latency and high-speed convolutions and matrix multiplication in optical domain; (3) }
%350 words max
\label{fig:intro}
\end{center}
\end{figure}

\section{Results} \label{Sec/res}

%Up to three levels of \textbf{subheading} are permitted. Subheadings should not be numbered.

Here we propose an optical compression/hashing system inspired by the SWIFFT family of post-quantum lattice-based hashing and encryption algorithms. This 4f-based free-space optical system accomplishes a two-step hashing and compression via matrix-matrix multiplication and convolution in Fourier domain. The samples of input versus output data from the algorithm, and the schematics of the proposed experimental setup is shown in Fig. \ref{fig:intro}. The envisioned optical setup consists of (a) a source, e.g. input delivered via fiber optics interconnect or an otherwise spatially modulated collimated laser beam; (b) SLM1 with a projected weight matrix $\omega_{mn}$ modulating in amplitude-only regime; (c) complex modulation unit comprising two SLMs embedded in a Michelson-type setup. The output of this system will have a 1D-like appearance and spread longitudinally. The cylindrical lens positioned in front of the detector array DA accomplishes the final step of the compression. The sensor's readout can subsequently be stored in a hash table and/or used as an input for a heterogeneous CNN classifier. Further technical details and the theoretical background are presented in Sec.\ref{Sec/methods}.

\begin{figure}[t!]
\begin{center}
\includegraphics[width=\linewidth]{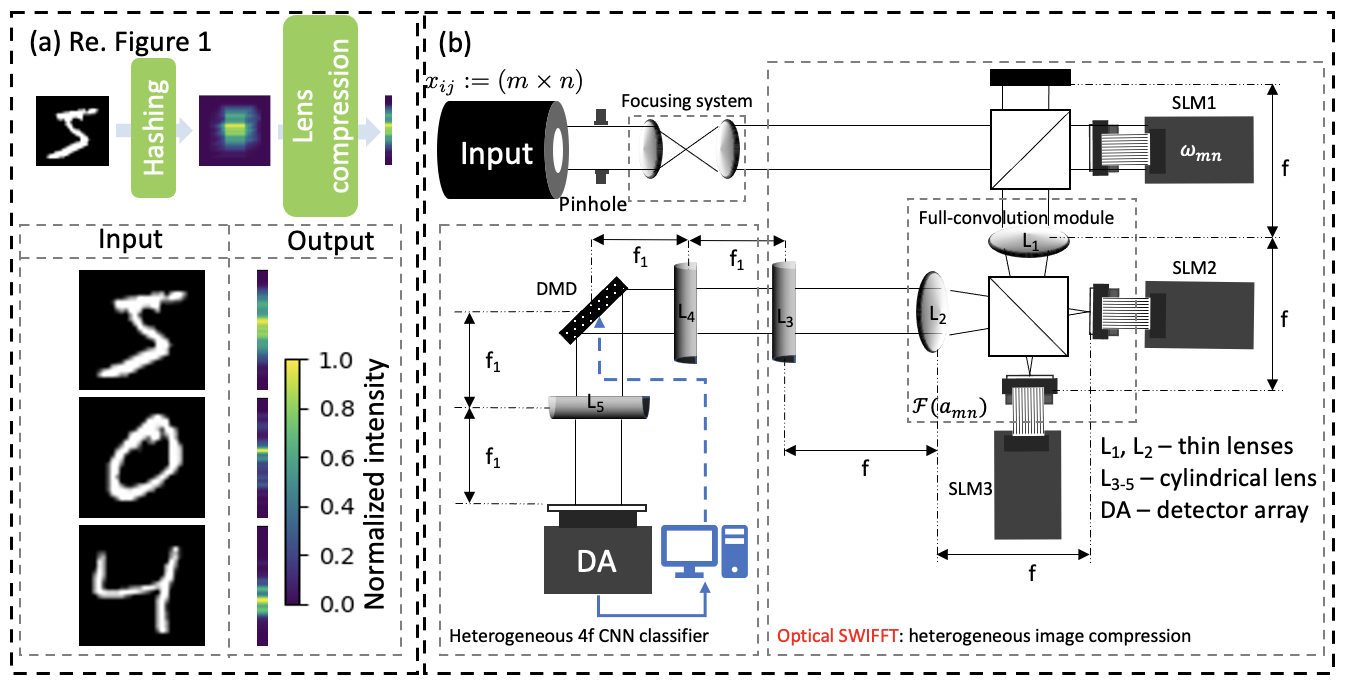}
\caption{The concept of optical hashing; (a) an example of an 1D compressed sequence, read of the simulated camera reading that corresponds to MNIST input "5", "0" and "4"; (b) the scheme of a proposed optical setup where source can be a single-mode fiber array collimated into free space or coupled in a photonic chip; diffusion can be achieved by an optical amplitude-only modulation by an SLM; and confusion must be done with complex  amplitude-and-phase convolution in the Fourier domain; the optional final step of image classification with a heterogeneous convolutional 4f classifier can be alternatively replaces by a CMOS diode array DA. \cite{miscuglio2020massively}.}
%350 words max
\label{fig:concept}
\end{center}
\end{figure}

The optical free-space propagation part of the featured compression algorithm was simulated electronically with the account of diffraction optics effects and within the paraxial approximation \cite{goodman2005introduction}. The simulated parameters of the optical setup in Fig. \ref{fig:concept}(b) are $\lambda = 633$nm is the light wavelength; the system's focal distance of the lenses $L_1$ and $L_2$ depends on the dimensions of the input mask as $f = \text{(mask size, nm)} \times \text{(pixel size)}/ \text{(wavelength, nm)}$. SLMs are assumed to have $2\pi$ pixel-to-pixel phase modulation range, 1910X1080 pixel resolution, and 8 bit pixel depth (e.g. PLUTO-2 from HOLOEYE).

\begin{figure}[t!]
\begin{center}
\includegraphics[width=\linewidth]{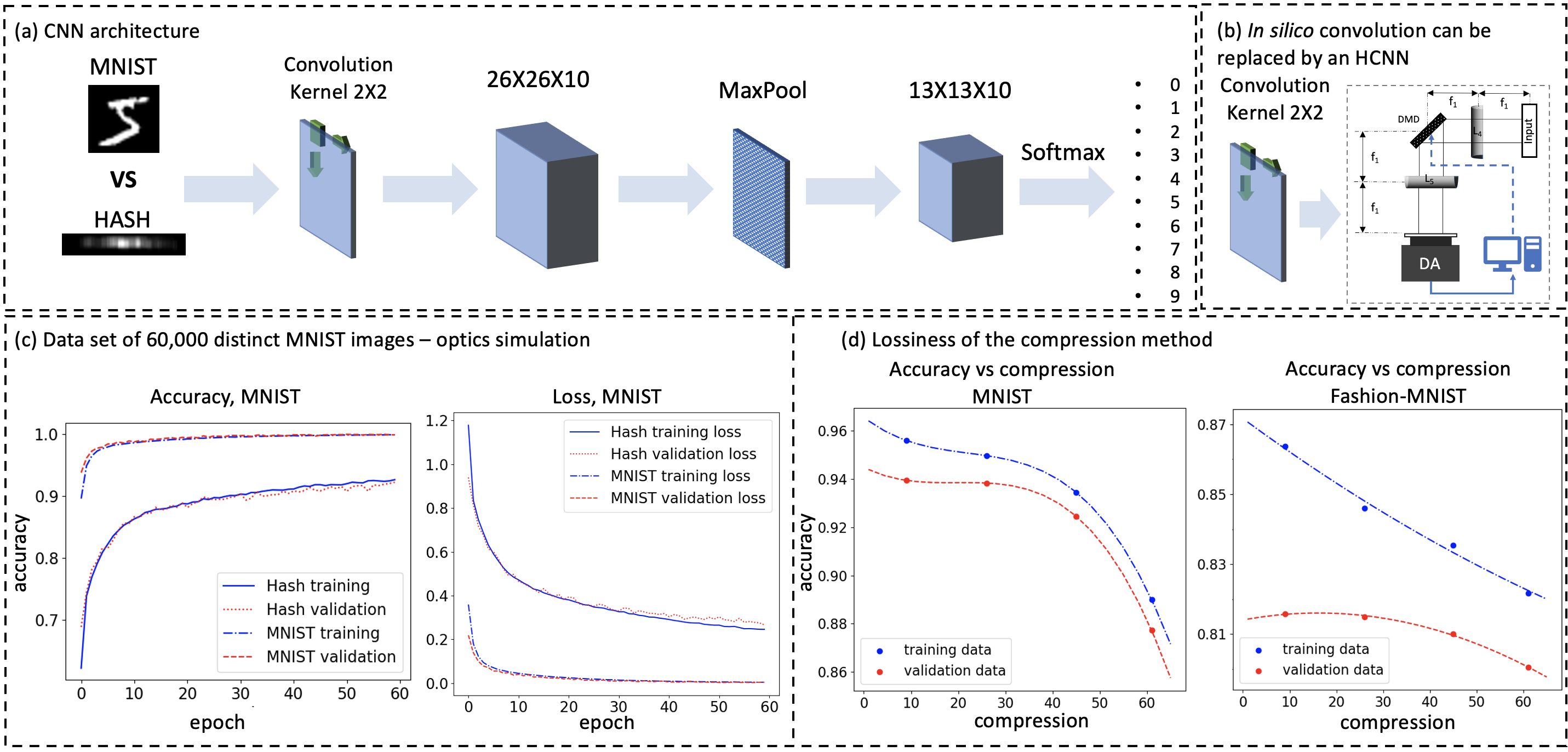}
\caption{CNN architecture and training results for MNIST compression/hashing algorithm; (a) the used in the experiment CNN architecture comprising one convolution, one pooling and one fully-connected layer with softmax activation function; (b) the convolution layer at training and inference stages are proposed to be replaced by an optical 4f-convolution layer \cite{miscuglio2020massively}; (c) training metric (accuracy and loss) for the original MNIST dataset of 60,000 images training part and the corresponding hashed/compressed images; the 60,000 images are separated into 40,000 for training and 20,000 for validation; (d) study of lossiness of the hashing/compression function, \textbf{Algorithm} \ref{algorithm_optical}, where information loss exhibits power-low behavior as a function of the compression coefficient.}
%350 words max
\label{fig:MNIST_results}
\end{center}
\end{figure}

The 60,000 hashed images generated from the training portion of MNIST dataset have been split and 40,000 have been used for training, and 20,000 classification in a 3-layer CNN: 2D convolution layer with 10 filters of size 3X3, maxpooling layer with $p=2$ pooling parameter, and the dense layer to output 10 categories for numbers from 0 to 9, see \textbf{Figure} \ref{fig:MNIST_results}(a). Training has been performed using Keras library in Python environment. A new dataset of 60,000 hashed images has been generated using our compression algorithm and utilizing the training set of MNIST is the input. Both original and hash datasets were trained on the same model with the same set of parameters: 60 epochs of batch size 1. It is worth to mention, that the choice of these parameters as well as the CNN architecture did not matter for the experiment. The goal is not to show the performance of a particular ML model on the given hashed data set in absolute, but to juxtapose performance of any given CNN on original MNIST data versus the hashed MNIST data.

The training and validation curves are shown in \textbf{Figure} \ref{fig:MNIST_results}(c), where the performance of the CNN model in \textbf{{Figure}} \ref{fig:MNIST_results}(a) on the original MNIST data is juxtaposed with its performance on generated in optical simulation dataset of hash/compressed MNIST images using the \textbf{Algorithm} \ref{algorithm_optical}. The results show that the proposed compression method successfully picks up features of the original images showing $\sim$8\% performance trade-off for compression 61 for 60k hash dataset generated in the optical simulation. We would like to emphasize, that this gap could be mitigated by increasing the depth of the NN in optical \cite{lin2018all} or electronic domain by increasing the number of parameters (layers) in the network architecture since systematic optical errors can be incorporated into the weights of the model by regular NN training techniques.

In the next step, we study classification accuracy as a function of the compression coefficient: (input resolution)/(output resolution). We start from compression of 10 and go up to 60 while keeping the aspect ratio $\sim$ 9:88. The compression is accomplished by averaging over the adjacent pixels to emulate the performance of cylindrical lens $L_3$ in \textbf{Figure} \ref{fig:concept}(b). The MNIST and Fashion-MNIST datasets have been considered. For Fashion-MNIST, we increased the depth of the NN model adding one more dense layer with 70 categories and relu activation to accomodate the higher complexity of the dataset with respect to MNIST. The performance trade-off is about 15\% for the compression of 60. The accuracy roll-off shows a similar tendency for both data set hashing/compression, see \textbf{Figure} \ref{fig:MNIST_results}(d). Based on these results, we project that compression loss may increase as a function of input data complexity due to the diffractive losses in an optical setup.

One last test has been performed on GTSRB \cite{GTSRB} dataset of German traffic signs with 50,000 RGB images in total and 43 classes. The images have been first converted into single channel grayscale, then the dataset has been augmented with additional images to insure equal number of images per class. The CNN has been expanded to accommodate more complex data to two convolution and two fully-connected layers with about 8,000,000 parameters. The training was terminated by early stopping under the condition of loss not changing for 50 epochs. With compression coefficient of 17 the validation accuracy loss is 54.4\%. This is a cumulative loss as the images have been both compressed \emph{and} converted to grayscale from RGB.

\begin{table}[ht]
\centering
\begin{tabular}{|l|l|l|l|}
\hline
Dataset & Compression coefficient & Drop, training (\%) & Drop, validation (\%) \\
\hline
MNIST & 60 & 10.4 & 11.9 \\
\hline
Fashion-MNIST & 60 & 18.2 & 19.2\\
\hline
GTSRB & 17 & 47.6 & 54.4\\
\hline
\end{tabular}
\caption{\label{tab:example}Table of accuracy trade-off for three datasets: MNIST (binary), Fashion-MNIST (binary) and GTSRB (RGB) binarized.}
\label{table_1}
\end{table}

Here, the \emph{in silico} CNN that is used for image classification, \ref{fig:MNIST_results}(a). However, it can be replaced by a heterogeneous 4f CNN (HCNN) classifier \cite{miscuglio2021massively, hu2021batch}, as shown in \textbf{Figure} \ref{fig:MNIST_results}(b) and in the scheme \textbf{Figure} \ref{fig:concept}(b). Delayed domain crossing can offer a benefit of harvesting maximum of the incoming analogue signal information and high-speed and high energy efficiency optical convolution layer, while not adding the optical IO related latency since the data is pre-existent in optical domain as it is.

%The proposed here optical data hashing method reduces the output image resolution by compressing the features of a 2D image into a 1D array that makes it ML-friendly for edge-computing purposes. However, it is worth mentioning, that this method is not lossless, as will be shown in the Results section of this paper: the $~$350 compression factor comes at a cost of $<1$\% classification accuracy in case with MNIST, ?\% for Fashion-MNIST, and ?\% for CIFAR10.

%\subsection*{Subsection}

%Example text under a subsection. Bulleted lists may be used where appropriate, e.g.

%\begin{itemize}
%\item First item
%\item Second item
%\end{itemize}

%\subsubsection*{Third-level section}
 
%Topical subheadings are allowed.

\section{Discussion}

We demonstrate a new opto-electronic hashing/compression algorithm capable of analogue 2D signal processing in optical domain with minimal domain conversions. The \textbf{Algorithm} \ref{algorithm_optical} is based on post-quantum hashing scheme SWIFFT \cite{lyubashevsky2008swifft}, see \textbf{Algorithm} \ref{Algorithm_electronic}, that takes advantage of information scrambling properties of Fourier transform and security of lattice-based algorithms. The proposed algorithm shows optical hardware-to-algorithm homomorphism as the associated tensor algebra operations, such as Fourier transform and by-element tensor multiplication, is native to Fresnel's optics and have shown the potential to be low-bandwidth and latency. Theoretical limits of this compression algorithm suggest a possibility of turning a input 2D-array into a 1D-array on the output.

As compared to other compression algorithms used in compressive sensing and digital compression, where information gets compressed before transmission or processing, and subsequently decompressed to recover the original image, we propose to use this algorithm in combination with heterogeneous NN training techniques. This way, hashed information is inputted into the heterogeneous ML accelerator where training and classification happens on already augmented data. The output of this this system, \emph{Figure} \ref{fig:concept}(b), can be used as a trigger for larger (and slower) edge data acquisition systems when the original high-resolution data of an event of interest is needed. Alternatively, for systems with small input data variability, look-up tables can be created to accomplish de-hashing back-end. The appeal is to reduce the requirement on throughput and latency of heterogeneous systems by pre-hashing and pre-compressing the data optically.

Our simulation results show, for compression coefficient 60 the MNIST accuracy drop is about 10.8\% and for Fashion-MNIST 18.5\%, see \textbf{Table} \ref{table_1}. For GTSRB 8 bit RGB dataset converted into 1-bit grayscale beforehand, the loss is $\sim$ 50\%. This is expected since the original digital version of SWIFFT algorithm is not lossless. We would like to mention, that when designing the corresponding ML networks we used a minimalistic approach keeping the architecture of the classifier simple. We believe, that performance of an ML algorithm on the compressed datasets can be boosted with more complex and tuned NN architectures. Our main  results and conclusions are: (1) the proposed algorithm is highly-parallelisable and capable of accomplishing fast and efficient 2D data compression with quadratically increasing loss as a function of compression coefficient information; (2) the algorithm is robust with respect to the image data of various nature (MNIST, Fashion-MNIST, GTSRB); (3) as a hash function, the resulting dataset fundamentally has features necessary for applications in cryptography, such as asymptotic security and collision resistance \cite{lyubashevsky2008swifft, lyubashevsky2006generalized}. In addition, the projected possibility of outputting a 1D-array back end suggest on-chip implementations for the further ultra-fast photonic data processing, as opposed to dealing with the known \emph{in silico} interconnect bottlenecks in optical pre-processors.

\section{Methods}\label{Sec/methods}
%An ideal hash function is a one-directional function that maps a binary message of arbitrary length onto a string of fixed length using a public key -- cipher. Converting a message into ciphertext should be computationally trivial, while inferring the original message from encrypted one -- impossible in the context of the appropriate for the task security definition.

\subsubsection*{SWIFFT Hashing Algorithm}

In this paper, we focus on one of the FFT-based compression/hashing algorithms that belong to the family of lattice hashing algorithms. Since optical crypto is an interdisciplinary class of problems, lets begin with revisiting some fundamental concepts of number theory, cryptography and optical processing, that are essential for understanding of the SWIFFT algorithm \cite{lyubashevsky2008swifft}.

First, most deployed crypto algorithms heavily employs modular arithmetic of rings. In number theory a ring $\mathds{Z}$ can be broadly defined as a set of elements with defined operation of addition and multiplication. Addition is always commutative, while multiplication might be not. Hence, a commutative ring is any ring with commutative multiplication. By number of elements, a ring can be infinite (e.g. a ring of real numbers), and finite (e.g. a ring of Pauli matrices). The SWIFFT is requiring modular rings $\mathds{Z}_p$, also called in group theory "quotient groups", of a modulo $p$ - a prime number.

Other two concepts that arise in mathematical theory of cryptography are "confusion" and "diffusion". Confusion is an algorithm that puts each value of an input message in dependence on several (maximum possible) pieces of a key. The goal of confusion is to scramble the input to achieve "one-way" functionality of a given hash algorithm. "Diffusion" is an algorithm that reduces collision probability by ensuring that small alteration of a plain text results in an avalanche of change in chiphertext.

With these concepts in hand, let's proceed to the description of the original SWIFFT algorithm. Given a binary message to encode $\mathcal{M}$, choose a large integer $k$. Set $n = 2^k$, a prime number $p$ such that $2n=p-1$, and choose an integer $m \geqslant \log p$. Split the binary input message $\mathcal{M}$ into a set of $m$ $n$-bit vectors such that a matrix $x_{i,j}$ of dimensions $\{m,n\}$ is formed. Generate a weight matrix $\omega_{nm}$ out of $m$ replicas of a sequence of length $n$, and a key $a_{mn}$ such that it consists of $n$ replicas of a sequence of length $m$. The elements of these sequences can be randomly drawn from a set of elements on a commutative ring $\mathds{Z}_p$ or use a more sophisticated generation mechanism. SWIFFT hashing algorithm is laid out below.

\vspace{.5cm}
\begin{algorithm}[H] \label{Algorithm_electronic}
  \SetKwInOut{Input}{Input}
  \SetKwInOut{Output}{Output}

  \Input{A sequence of binary digits $\mathcal{M}$ of length $m \times n$}
  \Output{1D hash function $\tilde z_n = \sum_n z_{mn}$}
    Multiply an input matrix $x_{mn}$ to an integer matrix of weights $\omega_{mn}$ to form a lattice $\mathcal{L} (\omega_{n}) = \omega_{mn} x_{mn}$\;
    Accomplish diffusion of an input message by taking a DFT of the result of the previous step  $y_{mn}=\mathcal{F}[\mathcal{L} (\omega_n)]$ \label{step2}\;
    DFT the earlier introduced key $\mathcal{A}_{mn} = \mathcal{F}[a_{mn}]$ \label{step3}\;
    Accomplish confusion by multiplying the result of step \ref{step2} by the result of step \ref{step3}: $\mathcal{A}_{mn} y_{mn}$\;
    Inverse Fourier transform the result $z_{mn}=\mathcal{F} \left [\mathcal{A}_{mn} y_{mn}\right]$\;
    To accomplish the data compression, sum the resulting matrix over $m$: $\sum_n z_{mn} = \tilde z_n$\;
 \caption{Electronic SWIFFT hashing algorithm}
\end{algorithm}
\vspace{.5cm}

Hence, SWIFFT hash function can be defined via the operation of convolution
\begin{gather}
    h(x_{i,j}) = \mathcal{L} (x_{mn}) \ast \mathcal{A}_{mn}
    \label{hash_function}
\end{gather}
where $\ast$ denotes a convolution defined for 2D as:
\begin{gather*}
\mathcal{L}_{ij} \ast \mathcal{A}_{ij} = \sum_{t=-m}^{m} \sum_{\ell=-n}^{n} \mathcal{L}_{i-t, j-\ell} \; \mathcal{A}_{t, \ell}
\end{gather*}

\subsubsection*{SWIFFT-based Optical Compression }

Let's first review several fundamental concepts that come from optical processing. The one to start with is the two-dimensional Fourier transform of an optical signal that is defined in, e.g., \cite{goodman2005introduction} eqn. (2-1):
\begin{equation}
    \mathcal{F}\{g(x,y)\} = \iint_{-\infty}^{\infty} dxdy\; g(x,y) \exp[-2\pi i (f_x x + f_y y)]
\end{equation}
where $f_x$ and $f_y$ are the $x$ and $y$ components of the corresponding spatial frequency spectrum. Interestingly, an ideal thin lens is a Fourier transform engine supplied by nature: when placing the input transparency centered at a front focal plane of an ideal thin lens one receives a Fourier-transformed signal at a rear focal plane. Despite the benefits, one needs to remember, that an ideal thin lens is, of course, a theoretical model of a real thin lens performance, that only works approximately in practice. Also, optical systems, free-space and on-chip, are prone to noise that has a potential to irreversibly corrupt a message. Keeping this noise near the level of common SNR of modern telecommunication links is one of the performance targets of optical hashing.

One more missing piece is the convolution theorem, e.g. \cite{goodman2005introduction}:
\begin{equation}
    [g * h](x) = \int_{-\infty}^{\infty} d\psi \;g(\psi) h(x - \psi) = \mathcal F \left(g(x)\right) \cdot \mathcal F \left(h(x)\right)\textbf{}
    \label{convolution_theorem}
\end{equation}
where "$*$" denotes the convolution of signal $g(x)$ with a mask $h(x)$. One can see that applying the Fourier transform to the original functions $g(x)$ and $h(x)$ turns an elaborate multi-step matrix multiplication into a simple by-element product. 

In the original SWIFFT implementation, convolution is achieved by applying the convolution theorem that equates convolution with dot-product multiplication in the Fourier domain \eqref{convolution_theorem}. Fourier transform $\mathcal{F}(\bigodot)$ is done electronically, typically with a fast Fourier transform or a modular Fourier transform algorithm. The computational complexity of such implementation is $n \log p$, \cite{gyorfi2012high}, which is an achievement as opposed to $mn$-complexity of the brute force convolution. However, the Fourier transform remains the main consumer of computational resources in this algorithm, slowing down both key generation and encryption/decryption. These result in lower key update rate and, consequently, compromises the security.

In this paper we propose, simulate and analyze an optical SWIFFT enspired data compression algorithm applied to benchmark ML data sets such as MNIST. In the envisioned scheme, see Fig. \ref{fig:concept}(a), the computational engine consists of an optical Fourier transform accomplished with lenses L1 and L2, and a dot-product multiplication done by SLMs. The whole 4f-systems forms a complex amplitude-and-phase convolution in the Fourier domain.

One starts from aquiring a two-dimensional incoming message $\mathcal{M}$ into a matrix $x_{mn}$. One also generates $n$ elements on a ring $R_m$ to form the set of coefficients $\omega_n$. In case with using SLMs, one naturally works on a ring $\mathds{Z}/256\mathds{Z}$, in other words, mod $256$. The dot-product multiplication in step 2 of an algorithm can be done electronically or optically, depending on the original domain of the incoming message $\mathcal{M}$. It is worth mentioning that key $a_{mn}$ often also built of a polynomial on a ring, similar to the sequence $w_n$ forming matrix $\omega_{mn}$ in step 2 of the SWIFFT.

Steps 3-6, convolution \eqref{hash_function} in Fourier domain, can be achieved optically using a 4f-system such as \cite{miscuglio2020massively}. A thin lens, being a natural spatial-to-Fourier domain converter, provides low latency (limited by speed of light), and high parallelism passive FT-engine. Dot-product multiplication is pseudo-passive as multiplication occurs on the surface of a light-modulating device that does not require processing power. The matrices, generated in step 2 and step 4 loaded onto light modulators requiring electric input and initial data processing. The Fourier transform of the key, needed in step 4, can as well be accomplished optically. Last step of the algorithm can be accomplished by either summing over the 2D reading of the camera CCD or by replacing a 2D CCD with a 1D diode array with controllable number of activated diodes, allowing parsing through the output 2D signal during the image collection time.

\vspace{.5cm}
\begin{algorithm}[H] \label{algorithm_optical}
  \SetKwInOut{Input}{Input}
  \SetKwInOut{Output}{Output}

  \Input{An incoming signal $\mathcal{M}$ of dimensions $m \times n$}
  \Output{CCD reading of hash matrix $|z_{mn}|^2$}
 Optically multiply $x_{mn}$ to an integer matrix of weights $\omega_{mn}$ by using modulating devices M1 and M2 in Fig. \ref{fig:concept}\;
 Optically Fourier transform the resulting signal to achieve diffusion $y_{mn}=\mathcal{F}[\mathcal{L} (\omega_n)]$\;
 Optically generate or electronically pre-calculate and project onto a modulating device M3 in Fig. \ref{fig:concept} the matrix $\mathcal{A}_{mn}=\mathcal{F}[a_{mn}]$ where it will get by-pixel multiplied by the output of step 2\;
 Fourier transform with L2: $z_{mn}=\mathcal{F} \left [\mathcal{A}_{mn} y_{mn}\right]$\;
 Compress with cylindrical lens L3\;
 Get the CCOS diode array reading $|z_{m}|^2$\;
 \caption{SWIFFT-like optical data compression algorithm}
\end{algorithm}
\vspace{.5cm}

For completeness, one could alternatively accomplish data compression at the domain crossing stage in step 5 by the use of reconfigurable diode arrays or electronically 

We have taken MNIST, Fashion-MNIST and GTSRB datasets as input data. The resulting hashed images for MNIST are shown in Fig. \ref{fig:concept}(b). The cipher images have been generated for the same diffusion matrix $\omega_{mn}$, and the same key $a_{mn}$. For additional details  see Sec. \ref{Sec/res}.

It is worth re-emphasizing that an accurate optical confusion can only be achieved if the dot-product multiplication in the Fourier domain captures both phase \emph{and} amplitude components. In particular, $\mathcal{A}_{mn}$ is a complex valued matrix with, generally speaking, non-trivial real and imaginary parts. The signal $x_{mn}$ will also carry a phase component after passing lens L1. Such complex convolution can be done interferometrically or with the help of new-generation modulating devices that are yet to be achieved experimentally.

\subsubsection*{Analysis}

An electronic simulation of the optical performance of the proposed hashing system has been done using Fresnel's transfer function as a propagator, and focusing function to simulate the lensing effect, see \cite{david2011computational}. It is important to satisfy the applicability criteria to obtain adequate results. In our simulation, the focal distance is related to the size of an incoming image, the size of the spatial modulator pixel, and the wavelength as $f = L \cdot \Delta/\lambda$. The input versus output hash are shown in Fig. \ref{fig:concept}(a). To correctly model such an optical setup in a simulation, just the same as in DFT, optical FT simulation needs padding. These results into an estimated input field of an image of about 250X250 pixel as opposed to its original size (e.g. 28X28 for MNIST and Fashion-MNIST). If the latter dimensions are taken as an input, then the compression rate of this system is $~60$. We would like to remind the readers, that comparing the nominal MNIST resolution to the system output resolution to assess compression efficiency in this case is not justified. For many potential applications, the input is analogue and the analogue-to-digital conversion is expected to happen only after the pre-processing with our custom system. Hence, direct comparison of the featured here information compression to the digital analogues is not straightforward. The presented system is not intended to replace any of the existing electronic systems. Instead, it is meant for use, e.g., in an edge processing setup for \emph{optical} pre-processing. Hence, the input is envisioned to be, e.g., an optical fiber array or a real-life scene that is usually captured by a several megapixels CCD/CMOS camera. Taking these into consideration, the cited compression coefficient is significant and can get even higher depending on the nature of the data and the demand on classification accuracy.

\bibliography{main}

\begin{thebibliography}{10}
\urlstyle{rm}
\expandafter\ifx\csname url\endcsname\relax
  \def\url#1{\texttt{#1}}\fi
\expandafter\ifx\csname urlprefix\endcsname\relax\def\urlprefix{URL }\fi
\expandafter\ifx\csname doiprefix\endcsname\relax\def\doiprefix{DOI: }\fi
\providecommand{\bibinfo}[2]{#2}
\providecommand{\eprint}[2][]{\url{#2}}

\bibitem{naughton2003efficient}
\bibinfo{author}{Naughton, T.~J.}, \bibinfo{author}{McDonald, J.~B.} \&
  \bibinfo{author}{Javidi, B.}
\newblock \bibinfo{journal}{\bibinfo{title}{Efficient compression of fresnel
  fields for internet transmission of three-dimensional images}}.
\newblock {\emph{\JournalTitle{Applied Optics}}} \textbf{\bibinfo{volume}{42}},
  \bibinfo{pages}{4758--4764} (\bibinfo{year}{2003}).

\bibitem{dufaux2015compression}
\bibinfo{author}{Dufaux, F.}, \bibinfo{author}{Xing, Y.},
  \bibinfo{author}{Pesquet-Popescu, B.} \& \bibinfo{author}{Schelkens, P.}
\newblock \bibinfo{journal}{\bibinfo{title}{Compression of digital holographic
  data: an overview}}.
\newblock {\emph{\JournalTitle{Applications of Digital Image Processing
  XXXVIII}}} \textbf{\bibinfo{volume}{9599}}, \bibinfo{pages}{163--173}
  (\bibinfo{year}{2015}).

\bibitem{zhang2020optical}
\bibinfo{author}{Zhang, L.}, \bibinfo{author}{Xiong, R.},
  \bibinfo{author}{Chen, J.} \& \bibinfo{author}{Zhang, D.}
\newblock \bibinfo{journal}{\bibinfo{title}{Optical image compression and
  encryption transmission-based on deep learning and ghost imaging}}.
\newblock {\emph{\JournalTitle{Applied Physics B}}}
  \textbf{\bibinfo{volume}{126}}, \bibinfo{pages}{1--10}
  (\bibinfo{year}{2020}).

\bibitem{donoho2006compressed}
\bibinfo{author}{Donoho, D.~L.}
\newblock \bibinfo{journal}{\bibinfo{title}{Compressed sensing}}.
\newblock {\emph{\JournalTitle{IEEE Transactions on information theory}}}
  \textbf{\bibinfo{volume}{52}}, \bibinfo{pages}{1289--1306}
  (\bibinfo{year}{2006}).

\bibitem{qaisar2013compressive}
\bibinfo{author}{Qaisar, S.}, \bibinfo{author}{Bilal, R.~M.},
  \bibinfo{author}{Iqbal, W.}, \bibinfo{author}{Naureen, M.} \&
  \bibinfo{author}{Lee, S.}
\newblock \bibinfo{journal}{\bibinfo{title}{Compressive sensing: From theory to
  applications, a survey}}.
\newblock {\emph{\JournalTitle{Journal of Communications and networks}}}
  \textbf{\bibinfo{volume}{15}}, \bibinfo{pages}{443--456}
  (\bibinfo{year}{2013}).

\bibitem{li2016performance}
\bibinfo{author}{Li, X.}, \bibinfo{author}{Zhang, G.}, \bibinfo{author}{Huang,
  H.~H.}, \bibinfo{author}{Wang, Z.} \& \bibinfo{author}{Zheng, W.}
\newblock \bibinfo{title}{Performance analysis of {GPU}-based convolutional
  neural networks}.
\newblock In \emph{\bibinfo{booktitle}{2016 45th International conference on
  parallel processing (ICPP)}}, \bibinfo{pages}{67--76}
  (\bibinfo{organization}{IEEE}, \bibinfo{year}{2016}).

\bibitem{jorda2019performance}
\bibinfo{author}{Jorda, M.}, \bibinfo{author}{Valero-Lara, P.} \&
  \bibinfo{author}{Pena, A.~J.}
\newblock \bibinfo{journal}{\bibinfo{title}{Performance evaluation of cudnn
  convolution algorithms on nvidia volta gpus}}.
\newblock {\emph{\JournalTitle{IEEE Access}}} \textbf{\bibinfo{volume}{7}},
  \bibinfo{pages}{70461--70473} (\bibinfo{year}{2019}).

\bibitem{naughton2002compression}
\bibinfo{author}{Naughton, T.~J.}, \bibinfo{author}{Frauel, Y.},
  \bibinfo{author}{Javidi, B.} \& \bibinfo{author}{Tajahuerce, E.}
\newblock \bibinfo{journal}{\bibinfo{title}{Compression of digital holograms
  for three-dimensional object reconstruction and recognition}}.
\newblock {\emph{\JournalTitle{Applied optics}}} \textbf{\bibinfo{volume}{41}},
  \bibinfo{pages}{4124--4132} (\bibinfo{year}{2002}).

\bibitem{mills2005effects}
\bibinfo{author}{Mills, G.~A.} \& \bibinfo{author}{Yamaguchi, I.}
\newblock \bibinfo{journal}{\bibinfo{title}{Effects of quantization in
  phase-shifting digital holography}}.
\newblock {\emph{\JournalTitle{Applied Optics}}} \textbf{\bibinfo{volume}{44}},
  \bibinfo{pages}{1216--1225} (\bibinfo{year}{2005}).

\bibitem{singh2010chaos}
\bibinfo{author}{Singh, N.} \& \bibinfo{author}{Sinha, A.}
\newblock \bibinfo{journal}{\bibinfo{title}{Chaos-based secure communication
  system using logistic map}}.
\newblock {\emph{\JournalTitle{Optics and Lasers in Engineering}}}
  \textbf{\bibinfo{volume}{48}}, \bibinfo{pages}{398--404}
  (\bibinfo{year}{2010}).

\bibitem{juleang2021optical}
\bibinfo{author}{Juleang, P.} \& \bibinfo{author}{Mitatha, S.}
\newblock \bibinfo{title}{Optical hash function for high speed and high
  security algorithm using ring resonator system}.
\newblock In \emph{\bibinfo{booktitle}{2021 7th International Conference on
  Engineering, Applied Sciences and Technology (ICEAST)}},
  \bibinfo{pages}{160--163} (\bibinfo{organization}{IEEE},
  \bibinfo{year}{2021}).

\bibitem{zhou2022photonic}
\bibinfo{author}{Zhou, H.} \emph{et~al.}
\newblock \bibinfo{journal}{\bibinfo{title}{Photonic matrix multiplication
  lights up photonic accelerator and beyond}}.
\newblock {\emph{\JournalTitle{Light: Science \& Applications}}}
  \textbf{\bibinfo{volume}{11}}, \bibinfo{pages}{1--21} (\bibinfo{year}{2022}).

\bibitem{zhou2019optical}
\bibinfo{author}{Zhou, J.} \emph{et~al.}
\newblock \bibinfo{journal}{\bibinfo{title}{Optical edge detection based on
  high-efficiency dielectric metasurface}}.
\newblock {\emph{\JournalTitle{Proceedings of the National Academy of
  Sciences}}} \textbf{\bibinfo{volume}{116}}, \bibinfo{pages}{11137--11140}
  (\bibinfo{year}{2019}).

\bibitem{situ2003cascaded}
\bibinfo{author}{Situ, G.} \& \bibinfo{author}{Zhang, J.}
\newblock \bibinfo{journal}{\bibinfo{title}{A cascaded iterative fourier
  transform algorithm for optical security applications}}.
\newblock {\emph{\JournalTitle{Optik}}} \textbf{\bibinfo{volume}{114}},
  \bibinfo{pages}{473--477} (\bibinfo{year}{2003}).

\bibitem{hennelly2003optical}
\bibinfo{author}{Hennelly, B.} \& \bibinfo{author}{Sheridan, J.~T.}
\newblock \bibinfo{journal}{\bibinfo{title}{Optical image encryption by random
  shifting in fractional fourier domains}}.
\newblock {\emph{\JournalTitle{Optics letters}}} \textbf{\bibinfo{volume}{28}},
  \bibinfo{pages}{269--271} (\bibinfo{year}{2003}).

\bibitem{situ2004double}
\bibinfo{author}{Situ, G.} \& \bibinfo{author}{Zhang, J.}
\newblock \bibinfo{journal}{\bibinfo{title}{Double random-phase encoding in the
  {F}resnel domain}}.
\newblock {\emph{\JournalTitle{Optics Letters}}} \textbf{\bibinfo{volume}{29}},
  \bibinfo{pages}{1584--1586} (\bibinfo{year}{2004}).

\bibitem{jin2007optical}
\bibinfo{author}{Jin, W.} \& \bibinfo{author}{Yan, C.}
\newblock \bibinfo{journal}{\bibinfo{title}{Optical image encryption based on
  multichannel fractional fourier transform and double random phase encoding
  technique}}.
\newblock {\emph{\JournalTitle{Optik}}} \textbf{\bibinfo{volume}{118}},
  \bibinfo{pages}{38--41} (\bibinfo{year}{2007}).

\bibitem{lang2010image}
\bibinfo{author}{Lang, J.}, \bibinfo{author}{Tao, R.} \& \bibinfo{author}{Wang,
  Y.}
\newblock \bibinfo{journal}{\bibinfo{title}{Image encryption based on the
  multiple-parameter discrete fractional fourier transform and chaos
  function}}.
\newblock {\emph{\JournalTitle{Optics Communications}}}
  \textbf{\bibinfo{volume}{283}}, \bibinfo{pages}{2092--2096}
  (\bibinfo{year}{2010}).

\bibitem{zhou2011novel}
\bibinfo{author}{Zhou, N.}, \bibinfo{author}{Wang, Y.} \&
  \bibinfo{author}{Gong, L.}
\newblock \bibinfo{journal}{\bibinfo{title}{Novel optical image encryption
  scheme based on fractional {M}ellin transform}}.
\newblock {\emph{\JournalTitle{Optics communications}}}
  \textbf{\bibinfo{volume}{284}}, \bibinfo{pages}{3234--3242}
  (\bibinfo{year}{2011}).

\bibitem{zhang2018multiple}
\bibinfo{author}{Zhang, L.}, \bibinfo{author}{Zhou, Y.}, \bibinfo{author}{Huo,
  D.}, \bibinfo{author}{Li, J.} \& \bibinfo{author}{Zhou, X.}
\newblock \bibinfo{journal}{\bibinfo{title}{Multiple-image encryption based on
  double random phase encoding and compressive sensing by using a measurement
  array preprocessed with orthogonal-basis matrices}}.
\newblock {\emph{\JournalTitle{Optics \& Laser Technology}}}
  \textbf{\bibinfo{volume}{105}}, \bibinfo{pages}{162--170}
  (\bibinfo{year}{2018}).

\bibitem{goodman2005introduction}
\bibinfo{author}{Goodman, J.~W.}
\newblock \bibinfo{journal}{\bibinfo{title}{Introduction to fourier optics}}.
\newblock {\emph{\JournalTitle{Introduction to {F}ourier optics, 3rd ed., by JW
  Goodman. Englewood, CO: Roberts \& Co. Publishers, 2005}}}
  \textbf{\bibinfo{volume}{1}} (\bibinfo{year}{2005}).

\bibitem{ajtai1996generating}
\bibinfo{author}{Ajtai, M.}
\newblock \bibinfo{title}{Generating hard instances of lattice problems}.
\newblock In \emph{\bibinfo{booktitle}{Proceedings of the twenty-eighth annual
  ACM symposium on Theory of computing}}, \bibinfo{pages}{99--108}
  (\bibinfo{year}{1996}).

\bibitem{lyubashevsky2008swifft}
\bibinfo{author}{Lyubashevsky, V.}, \bibinfo{author}{Micciancio, D.},
  \bibinfo{author}{Peikert, C.} \& \bibinfo{author}{Rosen, A.}
\newblock \bibinfo{title}{{SWIFFT}: A modest proposal for fft hashing}.
\newblock In \emph{\bibinfo{booktitle}{International workshop on fast software
  encryption}}, \bibinfo{pages}{54--72} (\bibinfo{organization}{Springer},
  \bibinfo{year}{2008}).

\bibitem{miscuglio2021massively}
\bibinfo{author}{Miscuglio, M.} \emph{et~al.}
\newblock \bibinfo{title}{Massively-parallel amplitude-only fourier optical
  convolutional neural network}.
\newblock In \emph{\bibinfo{booktitle}{2021 Conference on Lasers and
  Electro-Optics (CLEO)}}, \bibinfo{pages}{1--2} (\bibinfo{organization}{IEEE},
  \bibinfo{year}{2021}).

\bibitem{miscuglio2020photonic}
\bibinfo{author}{Miscuglio, M.} \& \bibinfo{author}{Sorger, V.~J.}
\newblock \bibinfo{journal}{\bibinfo{title}{Photonic tensor cores for machine
  learning}}.
\newblock {\emph{\JournalTitle{Applied Physics Reviews}}}
  \textbf{\bibinfo{volume}{7}}, \bibinfo{pages}{031404} (\bibinfo{year}{2020}).

\bibitem{miscuglio2020massively}
\bibinfo{author}{Miscuglio, M.} \emph{et~al.}
\newblock \bibinfo{journal}{\bibinfo{title}{Massively parallel amplitude-only
  fourier neural network}}.
\newblock {\emph{\JournalTitle{Optica}}} \textbf{\bibinfo{volume}{7}},
  \bibinfo{pages}{1812--1819} (\bibinfo{year}{2020}).

\bibitem{lin2018all}
\bibinfo{author}{Lin, X.} \emph{et~al.}
\newblock \bibinfo{journal}{\bibinfo{title}{All-optical machine learning using
  diffractive deep neural networks}}.
\newblock {\emph{\JournalTitle{Science}}} \textbf{\bibinfo{volume}{361}},
  \bibinfo{pages}{1004--1008} (\bibinfo{year}{2018}).

\bibitem{GTSRB}
\bibinfo{author}{Real-Time Computer Vision~group, I.}
\newblock \bibinfo{title}{{German Traffic Sign Recognition Benchmark}}.
\newblock
  \bibinfo{howpublished}{\url{https://www.kaggle.com/datasets/meowmeowmeowmeowmeow/gtsrb-german-traffic-sign}}
  (\bibinfo{year}{2011}).

\bibitem{hu2021batch}
\bibinfo{author}{Hu, Z.} \emph{et~al.}
\newblock \bibinfo{journal}{\bibinfo{title}{Batch processing and data streaming
  fourier-based convolutional neural network accelerator}}.
\newblock {\emph{\JournalTitle{arXiv preprint arXiv:2112.12297}}}
  (\bibinfo{year}{2021}).

\bibitem{lyubashevsky2006generalized}
\bibinfo{author}{Lyubashevsky, V.} \& \bibinfo{author}{Micciancio, D.}
\newblock \bibinfo{title}{Generalized compact knapsacks are collision
  resistant}.
\newblock In \emph{\bibinfo{booktitle}{International Colloquium on Automata,
  Languages, and Programming}}, \bibinfo{pages}{144--155}
  (\bibinfo{organization}{Springer}, \bibinfo{year}{2006}).

\bibitem{gyorfi2012high}
\bibinfo{author}{Gy{\"o}rfi, T.}, \bibinfo{author}{Cret, O.},
  \bibinfo{author}{Hanrot, G.} \& \bibinfo{author}{Brisebarre, N.}
\newblock \bibinfo{journal}{\bibinfo{title}{High-throughput hardware
  architecture for the swifft/swifftx hash functions.}}
\newblock {\emph{\JournalTitle{IACR Cryptol. ePrint Arch.}}}
  \textbf{\bibinfo{volume}{2012}}, \bibinfo{pages}{343} (\bibinfo{year}{2012}).

\bibitem{david2011computational}
\bibinfo{author}{David, V.}
\newblock \bibinfo{title}{Computational fourier optics} (\bibinfo{year}{2011}).

\end{thebibliography}

%\noindent LaTeX formats citations and references automatically using the bibliography records in your .bib file, which you can edit via the project menu. Use the cite command for an inline citation, e.g.  \cite{Hao:gidmaps:2014}.

%For data citations of datasets uploaded to e.g. \emph{figshare}, please use the \verb|howpublished| option in the bib entry to specify the platform and the link, as in the \verb|Hao:gidmaps:2014| example in the sample bibliography file.

\section{Acknowledgements}

%Acknowledgements should be brief, and should not include thanks to anonymous referees and editors, or effusive comments. Grant or contribution numbers may be acknowledged.

M.S.G acknowledges valuable discussions on algorithmic approaches and simulation details with Jonathan George and Nicholas Gorgone.

%\section{Author contributions statement}

%Must include all authors, identified by initials, for example:
%A.A. conceived the experiment(s),  A.A. and B.A. conducted the experiment(s), C.A. and D.A. analysed the results.  All authors reviewed the manuscript. 

%\section{Additional information}

%To include, in this order: \textbf{Accession codes} (where applicable); \textbf{Competing interests} (mandatory statement). 

%The corresponding author is responsible for submitting a \href{http://www.nature.com/srep/policies/index.html#competing}{competing interests statement} on behalf of all authors of the paper. This statement must be included in the submitted article file.

%\begin{table}[ht]
%\centering
%\begin{tabular}{|l|l|l|}
%\hline
%Condition & n & p \\
%\hline
%A & 5 & 0.1 \\
%\hline
%B & 10 & 0.01 \\
%\hline
%\end{tabular}
%\caption{\label{tab:example}Legend (350 words max). Example legend text.}
%\end{table}

%Figures and tables can be referenced in LaTeX using the ref command, e.g. Figure \ref{fig:stream} and Table \ref{tab:example}.

\end{document}